\documentclass[aps,prd,nofootinbib,superscriptaddress,preprintnumbers]{revtex4}
\usepackage{rotating}
\usepackage{amssymb}
\usepackage{amsmath}
\usepackage{epsfig}
\usepackage{hyperref}
\usepackage{color}
\usepackage{graphicx}

\makeatletter
\def\simgt{\mathrel{\lower2.5pt\vbox{\lineskip=0pt\baselineskip=0pt
           \hbox{$>$}\hbox{$\sim$}}}}
\def\simlt{\mathrel{\lower2.5pt\vbox{\lineskip=0pt\baselineskip=0pt
           \hbox{$<$}\hbox{$\sim$}}}}
\makeatother

\newcommand{\cev}[1]{\reflectbox{\ensuremath{\vec{\reflectbox{\ensuremath{#1}}}}}}
\newcommand{\be}{\begin{equation}}
\newcommand{\ee}{\end{equation}}
\newcommand{\bea}{\begin{eqnarray}}
\newcommand{\eea}{\end{eqnarray}}
\newcommand{\eq}[2]{\be\begin{aligned}#1 \label{#2}\end{aligned}\ee}

\newcommand{\tabeq}[2]{ \parbox{#1}{  \be\begin{aligned}#2 \end{aligned} \nonumber \ee }}

\newcommand{\Fig}[1]{Fig.~\ref{#1}}
\newcommand{\Eq}[1]{Eq.~\eqref{#1}}

\newcommand{\Sec}[1]{Sec.~\ref{#1}}

\begin{document}

\title{Mining the Geodesic Equation for Scattering Data}

\author{Clifford Cheung}
\affiliation{Walter Burke Institute for Theoretical Physics,
California Institute of Technology, Pasadena, CA 91125}

\author{Nabha Shah}
\affiliation{Walter Burke Institute for Theoretical Physics,
California Institute of Technology, Pasadena, CA 91125}

\author{Mikhail P. Solon}
\affiliation{Mani L. Bhaumik Institute for Theoretical Physics,
Department of Physics and Astronomy, UCLA, Los Angeles, CA 90095}

\begin{abstract}

The geodesic equation encodes test-particle dynamics at arbitrary gravitational coupling, hence retaining all orders in the post-Minkowskian (PM) expansion. Here we explore what geodesic motion can tell us about dynamical scattering in the presence of perturbatively small effects such as tidal distortion and higher derivative corrections to general relativity. We derive an algebraic map between the perturbed geodesic equation and the leading PM scattering amplitude at arbitrary mass ratio. As examples, we  compute formulas for amplitudes and isotropic gauge Hamiltonians for certain infinite classes of tidal operators such as electric or magnetic Weyl to any power, and for higher derivative corrections to gravitationally interacting bodies with or without electric charge. Finally, we present a general method for calculating closed-form expressions for amplitudes and isotropic gauge Hamiltonians in the test-particle limit at all orders in the PM expansion.

\end{abstract}

\preprint{CALT-TH 2020-042}

\maketitle


\section{Introduction}

In recent years, powerful tools from the modern amplitudes program~\cite{DoubleCopy,GeneralizedUnitarity} and effective field theory have~\cite{oldQFT,NRGR,classical} been unified to derive new results~\cite{2PM,3PM,3PMlong,3PMFeynman} of relevance to the search for gravitational waves at the LIGO/Virgo experiment~\cite{LIGO}. These efforts have spurred a resurgence of interest in post-Minkowskian (PM) perturbation theory~\cite{PM}, which organizes dynamics in powers of the gravitational constant $G$ while retaining all orders in velocity.  This approach lies in contrast with post-Newtonian (PN) perturbation theory \cite{PN}, which further expands in velocity as appropriate for a virialized system.  The PM and PN formalisms, together with methods from numerical relativity~\cite{NR}, effective one-body theory~\cite{EOB}, self-force~\cite{self_force}, and effective field theory~\cite{NRGR,NRGR2} form a vital ecosystem of ideas and tools for solving the binary inspiral problem.  Many recent developments have benefited immensely from a genuine cross-pollination of ideas between classical general relativity and quantum field theory~\cite{newAMP,HBET,N8,GravitonScattering}, culminating in new results pertaining to spinning black holes~\cite{newSPIN,Vines:2018gqi,Zvispin}, orbital mechanics~\cite{LSE, B2B1, Bjerrum-Bohr:2019kec, B2B2}, and radiation~\cite{radiation}.

The preponderance of work in this area has centered on black holes in general relativity.  There are, however, many reasons to consider perturbations away from this minimal scenario.    A prime example of this is tidal distortion, which is central to disentangling the inspiral dynamics of neutron binary systems that will hopefully shed light on the underlying nuclear properties of dense matter \cite{Buonanno:2014aza,Dietrich:2020eud}. Indeed, gravitational waves from neutron binaries should place substantive constraints on the nuclear equation of state~\cite{TheLIGOScientific:2017qsa,Abbott:2020uma,Flanagan:2007ix, MeasuringTidal}.  For these reasons, tidal phenomena have been modeled with a variety of methods~\cite{NRTidal,Flanagan:2007ix, AnalyticTidal,Steinhoff:2016rfi,Henry:2019xhg,SFTidal,EOBTidal}, including, very recently, PM perturbation theory~\cite{Bini:2020flp,PMGR,PMtidal,Goldberger:2020wbx,Haddad:2020que,Kalin:2020lmz}. 

Deviations from the minimal scenario also arise in theories of modified gravity.  In the absence of additional light degrees of freedom such dynamics are encoded at long distances by higher derivative corrections to general relativity~\cite{Endlich:2017tqa,Cardoso:2018ptl,Sathyaprakash:2019yqt,Sennett:2019bpc}.   While these effects will be exceedingly small in any sensible ultraviolet completion of gravity it is still worth understanding their influence on the dynamics of the binary inspiral~\cite{Brandhuber:2019qpg,Emond:2019crr,Cristofoli:2019ewu,Cai:2019npx}.  

In this paper we explore how the geodesic equation encodes conservative dynamics in the presence of an arbitrary perturbative correction away from a non-spinning black hole binary system in general relativity.   We will be interested in computing the perturbed scattering amplitudes and Hamiltonians which characterize these effects.  Our main application will be tidal interactions.  However, given the generality of our approach we will also study other examples, including higher derivative corrections to Einstein-Maxwell theory and their effects on gravitationally and electrically interacting bodies.

As is well-known, the test-particle limit---and corrections away from it~\cite{BDG,BDG2,Antonelli:2020aeb}---encode critical information about scattering in the PM approximation~\cite{Vines:2018gqi,Damour:2019lcq}. More trivially, this limit completely fixes the dynamics of spinning and nonspinning black holes at low PM orders and offers a useful consistency check at higher orders~\cite{2PM, 3PM, 3PMlong, 3PMFeynman, PMtidal, Siemonsen:2019dsu, Zvispin}.  At a practical level, the test-particle regime is attractive since in many cases it is analytically tractable to all PM orders.   While this observation is somewhat trivial it is tantalizing in the context of scattering amplitudes, where it implies that the geodesic equation effectively resums certain infinite towers of loop diagrams.  An enticing possibility then emerges of bootstrapping certain multi-loop scattering calculations directly from geodesic motion.

Here we present two methods, each with its own advantages and target applications. The first applies at leading nontrivial PM order and linear order in some additional perturbative parameter.  Our method is based on the connection between PM Hamiltonians and scattering amplitudes~\cite{2PM}, and derives the tidal Hamiltonian in isotropic coordinates from the geodesic equation through a set of trivial algebraic operations. A key simplification is that at leading PM order, the canonical transformation from non-isotropic to isotropic coordinates is encapsulated by a replacement rule shown in \Eq{eq:FT}.   In the context of tidal interactions, this approach elaborates on the essential insight of~\cite{Bini:2020flp} that such contributions are encoded entirely by the geodesic dynamics of a tidally distorted test particle in a Schwarzschild background. We demonstrate the simplicity of our approach by deriving analytic formulas for scattering amplitudes and Hamiltonians at arbitrary mass ratio for certain infinite classes of tidal operators, including electric Weyl and magnetic Weyl to any power, as shown in \Eq{eq:H_En} and \Eq{eq:H_Bn}.  We also apply this method to other types of perturbative corrections, e.g.~which arise for electrically charged bodies or from higher derivative corrections to Einstein-Maxwell theory.
   
Our second method is a mechanical procedure for deriving closed-form expressions for the scattering amplitude in the test-particle limit at all orders in the PM expansion. We systematically perform a general diffeomorphism on the metric and then constrain it to eliminate non-isotropic terms in the geodesic equation. A general formula for the isotropic Hamiltonian induced by any tidal moment operator is presented in \Eq{eq:Hiso}. From this Hamiltonian we then derive expressions for the corresponding scattering amplitudes at all PM orders in \Eq{eq:Miso}.  As an application of these general formulas, we derive all orders in PM expressions for the isotropic Hamiltonian and scattering amplitude for a set of tidal operators, as presented in Table.~\ref{table:Hamiltonians} and Table.~\ref{table:amplitudes}.  These results may provide useful data for checking future higher order PM calculations.
\vspace{0.3cm}

\noindent {\it Note added:} During the completion of this project we learned of the interesting concurrent work of Ref.~\cite{Zvi_tidal}, which has similar results at leading PM order obtained via direct evaluation of multi-loop scattering amplitudes.   Their approach is nicely complementary to our own and where our results overlap they agree completely.  We thank the authors of Ref.~\cite{Zvi_tidal} for sharing their work with us before submission.  

\section{Test-Particle Dynamics}\label{sec:TPdynamics}

In this section we present some basic tools for deriving scattering amplitudes from the geodesic equation. For concreteness we describe this formalism in the context of tidal effects, but as we will see later on our essential approach is trivially generalized to any scenario in which the geodesic equation is perturbatively corrected.

To begin, we consider the scattering amplitude of a tidally distorted body of mass $m_1= m$ interacting with a black hole of mass $m_2 =M$ in the test-particle limit $m \ll M$.   In the rest frame of the black hole, the dynamics are described by the geodesic motion of a non-minimally coupled test particle of three-momentum $p$ propagating on a Schwarzschild background.
An arbitrary tidal moment is represented by the field theoretic operator,
\eq{
\Delta S =\frac{\lambda}{2} \int d^4 x\,  \phi \, {\cal O}(g,\cev{\nabla},\vec{\nabla}) \, \phi \,,
}{eq:action}
where ${\cal O}$ is some operator that involves the metric tensor $g$, the matter field $\phi$, and their derivatives.  We work at linear order in the tidal parameter $\lambda$ throughout.  

In order to make contact with known results, we express ${\cal O}$ in a basis of operators composed of products of the electric and magnetic Weyl tensors and their derivatives.  Note, however, that these tensors are defined with respect to the four-momentum of a propagating body and so one naively encounters an ambiguity in \Eq{eq:action}: should the four-momentum be associated with $\cev{\nabla}$ or $\vec{\nabla}$?  As it turns out, this choice is actually irrelevant since the difference is suppressed by the momentum transfer carried away by gravitons relative to the momentum of the bodies.   This suppression factor is infinitesmal, scaling inversely with the angular momentum of the scattering process in units of $\hbar$, as expected for a quantum correction to the dynamics \cite{2PM, 3PM, 3PMlong}. 

\subsection{Geodesic to Hamiltonian}

In the presence of tidal distortion the geodesic equation is \cite{Steinhoff:2016rfi} 
\eq{
0= g^{\mu\nu} P_\mu P_\nu + m^2 - \lambda {\cal O}(g, P) \,,
}{eq:geodesic}
in mostly plus signature and
where as discussed earlier we identify $\nabla$ with the four-momentum $P$ of the test particle. For maximum generality we assume an arbitrary static, spherically symmetric metric whose line element is 
\eq{
ds^2 = g_{tt}(r) dt^2 + g_{rr}(r) dr^2 + g_{\Omega}(r)  r^2 (d\theta^2 + \sin^2\theta \, d\phi^2) \,.
}{eq:metric}
This form of the metric will accommodate all our scenarios of interest, i.e.~black holes in general relativity and beyond in various coordinate systems.

The four-momentum of the test particle is
\eq{
P_\mu = (H, p_r, p_\theta, p_\phi) = \left(H, \sqrt{p^2 -\tfrac{J^2}{r^2}}, 0, J\right),
}{eq:P}
where $H$ is the test-particle Hamiltonian and $p_r = p \cdot r / r$ is the radial momentum.  In the last equality we have assumed a trajectory on the equatorial plane at $\theta= \pi/2$, so $p_\theta=0$ and $p_\phi= J$ is the angular momentum of the particle. We have also used $p_r^2 = p^2 - \frac{J^2}{r^2}$ to relate the radial momentum to the total momentum $p$ and angular momentum $J$.

To derive the Hamiltonian we first expand to linear order in the tidal coefficient, $H = H_0 + \lambda H_1 + {\cal O}(\lambda^2)$, and then solve the geodesic equation order by order in $\lambda$.  
 At zeroth order in the tidal coefficients, the Hamiltonian is 
\eq{
H_0(p,r,J) &=  \sqrt{-g_{tt}(r)} \, \sqrt{m^2+   \tfrac{p^2 - \frac{J^2}{r^2}}{g_{rr}(r)} + \tfrac{J^2}{g_\Omega(r)r^2}} \, .
}{eq:H0}
From \Eq{eq:geodesic} we see that the tidal operator manifestly enters as a shift of the mass term.  Thus in the presence of tidal distortion the Hamiltonian becomes
\eq{
H(p,r,J) &= \sqrt{-g_{tt}(r)} \, \sqrt{m^2 - \lambda {\cal O}(p,r,H_0, J)   + \tfrac{p^2 - \frac{J^2}{r^2}}{g_{rr}(r)} + \tfrac{J^2}{g_\Omega(r)r^2}}  + {\cal O}(\lambda^2)  \, .
}{eq:H_compact}
Let us elaborate on the appearance of ${\cal O}(p,r,H_0, J)$ in the above expression. A priori, the tidal operator is of the form ${\cal O}(g, P) = {\cal O}(p,r, H, J)$, which depends on the radius $r$ through the background metric $g$, and on $p$, $H$, and $J$ through the four-momentum $P$.   However, the tidal operator can be approximated by ${\cal O}(p,r,H,J)= {\cal O}(p,r,H_0,J) + {\cal O}(\lambda)$ up to corrections subleading in $\lambda$.  Further linearizing \Eq{eq:H_compact} in $\lambda$ we obtain the tidal Hamiltonian, 
\eq{
 H_1(p,r,J) &=  \frac{ g_{tt}(r)  }{ 2H_0} {\cal O}(p,r,H_0, J) \,.
}{eq:H1_LO}
While the focus of the present work is the leading correction in $\lambda$, it is of course straightforward to solve \Eq{eq:geodesic} at any arbitrary order in $\lambda$, yielding the corresponding tidal Hamiltonian at that order.

We will often be interested in corrections at the leading nontrivial PM order, in which case \Eq{eq:H1_LO} simplifies since $g_{tt}  \simeq -1$ and $H_0 \simeq E = \sqrt{p^2+ m^2}$ is the energy of a free particle, so
\eq{
H_1(p,r,J) = -\frac{{\cal O}(p,r,E,J)}{2 E} + \textrm{higher order in PM} \,.
}{eq:H1_simp} 
We thus conclude that the Hamiltonian at leading order in $\lambda$ and leading PM order is literally the tidal operator evaluated on the Newtonian metric with all momenta taken to be on-shell.

As we will see, it is often convenient to study the dynamics in isotropic gauge, i.e.~coordinates in which the Hamiltonian is a function $H^\textrm{iso}(p,r)$ which depends only on $p$ and $r$ but not $J$.  At zeroth order in tidal corrections, i.e.~for bodies with minimal gravitational coupling, $H_0^\textrm{iso}(p,r)$ is trivially obtained by choosing isotropic coordinates for the black hole metric,
\eq{
g_{tt}^{\rm iso}(r) = - \left( { 1- \frac{R}{4r}  \over 1 + \frac{R}{4r} }   \right)^2 \,, \qquad   g_{rr}^{\rm iso}(r) = g_{\Omega}^{\rm iso}(r) =  \left( { 1+ \frac{R}{4r}  }   \right)^4,
}{eq:iso_coord}
where $R = 2GM$ is the Schwarzschild radius.  Plugging \Eq{eq:iso_coord} into \Eq{eq:H0}, we obtain the isotropic Hamiltonian for a point-like test particle \cite{WS93},
\eq{
H_0^{\rm iso}(p,r) &=  \sqrt{-g_{tt}^{\rm iso}(r)} \sqrt{m^2 +\tfrac{p^2}{g_{rr}^{\rm iso}(r)}} \, ,
}{eq:Hiso0}
which we will use frequently for the remainder of the paper.

\subsection{Hamiltonian to Amplitude}\label{sec:HtoA}

Starting from an arbitrary Hamiltonian it is straightforward to compute the scattering amplitude for elastic scattering via effective field theory methods \cite{2PM}.  This approach requires the tedious albeit mechanical computation of loop diagrams order by order in the PM expansion.  On the other hand, for a Hamiltonian which is already in isotropic gauge there exists an extraordinarily simple procedure which bypasses loop integration in favor of solving an algebraic equation.  First discovered in the 3PM calculation of \cite{3PM, 3PMlong} and later proven in \cite{B2B1,Bjerrum-Bohr:2019kec}, this map exploits an elegant algebraic relation between the amplitude ${\cal M}$ and the isotropic gauge Hamiltonian $H^{\rm iso}$.  In the test-particle limit this relation is
\eq{
{\cal M}(p ,r) = \frac{1}{2E}\left({\bar p}(r)^2 - p^2\right)+\textrm{iterations} \, ,
}{eq:impetus}
where ${\bar p}(r)$ is the local momentum at a position $r$ dictated by the conservation of energy equation,
\eq{
H^\textrm{iso}({\bar p}(r),r)=E \,.
}{eq:energy_conservation}
The iteration contributions in \Eq{eq:impetus} are infrared divergent terms defined in the prescription of \cite{2PM}.  These contributions always cancel in any effective field theory matching to extract coefficients in the Hamiltonian.

Note that in the scattering amplitude  ${\cal M}(p,r)$ we should interpret the quantities $E=\sqrt{p^2+m^2}$ and $p$   as the asymptotic energy and momentum of the test particle and the variable $r$ as the Fourier transform of the three-momentum transfer $q$.   This differs from $H^{\rm iso}(p,r)$, where $p$ and $r$ should be interpreted as the time-dependent phase space coordinates of the test particle.  

The upshot here is that the isotropic Hamiltonian and scattering amplitude are trivially related.   In particular, we expand the scattering amplitude as ${\cal M} = {\cal M}_0 + \lambda {\cal M}_1 + {\cal O}(\lambda^2)$ and solve \Eq{eq:impetus} and \Eq{eq:energy_conservation} order by order in $\lambda$.  This procedure yields the tidally corrected scattering amplitude in the test-particle limit,
\eq{
 {\cal M}_0(p,r) &= \frac{1}{2E} \left[ m^2(1-g_{rr}^{\rm iso}(r)) - E^2\left(1+\tfrac{g_{rr}^{\rm iso}(r)}{g_{tt}^{\rm iso}(r)}\right) \right] + \textrm{iterations}  \\
{\cal M}_1(p,r) &= {g_{rr}^{\rm iso}(r) \over g_{tt}^{\rm iso}(r)} H_1^\textrm{iso}(\sqrt{E^2 + 2E {\cal M}_0 - m^2},r) + \textrm{iterations} \, ,
}{eq:Mgen}
where in the first line we have solved \Eq{eq:energy_conservation} at zeroth order in the tidal coefficient using $H_0^{\rm iso}$ from \Eq{eq:Hiso0}.  In the second line we have the leading tidal correction to the amplitude in terms of an abstract $H_1^{\rm iso}$ which we will compute explicitly later on.

At leading PM order, the tidal correction to the amplitude is
\eq{ 
{\cal M}_1(p,r) &= - H_1^\textrm{iso}(p,r) + \textrm{higher order in PM} \,,
}{eq:M1} 
so as expected the amplitude is exactly the Feynman vertex defined by the isotropic Hamiltonian.

\section{Leading Order in $G$}\label{sec:LO}

\begin{figure}[t]
\begin{center}
\includegraphics[scale=.48]{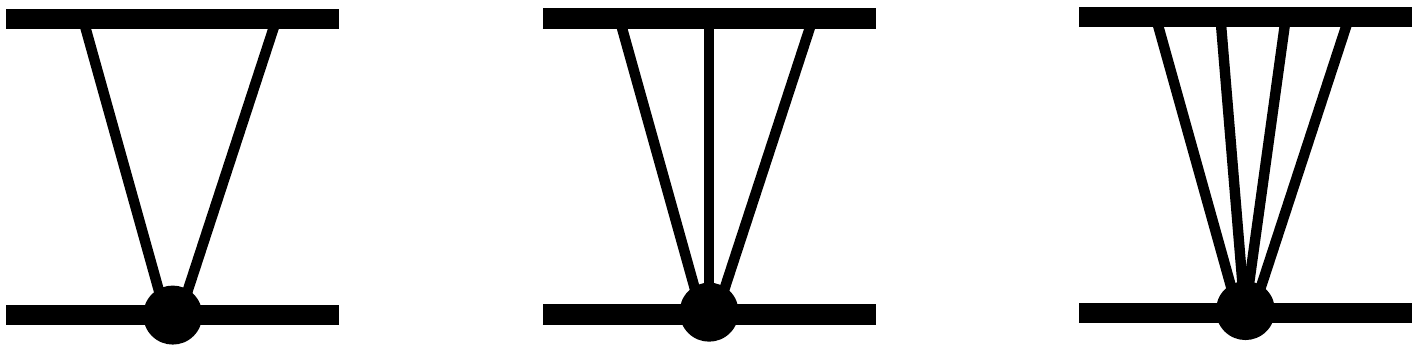}
\end{center}
\caption{At leading PM order, tidal corrections to the scattering amplitude arise from fan diagrams.  For an operator composed of $n$ curvature tensors, this contribution corresponds to an $(n-1)$-loop diagram. Thick and thin lines denote matter and graviton lines while the black dots denote tidal operator insertions.}
\label{fig:fan}
\end{figure}

In this section we consider the dynamics at leading PM order and leading order in some additional perturbative correction.  Our aim is to compute the perturbed scattering amplitude and Hamiltonian.   In such a regime the test-particle limit encodes complete information about the dynamics at arbitrary mass ratio.  This fact is obvious from the point of view of scattering amplitudes.  Consider, for example, tidal corrections at leading PM order.  These contributions are generated by the lowest order loop diagrams which induce classical scattering \cite{3PM, 3PMlong} and enter at linear order in the tidal coefficient, corresponding to the ``fan diagrams'' depicted in \Fig{fig:fan}.  By definition, fan diagrams do not include matter propagators of the tidally distorted particle.   More generally, we will henceforth refer to any diagram in which all matter propagators are on one side as a fan diagram.
As discussed in \cite{2PM,3PM,3PMlong,3PMFeynman}, all fan diagrams are free from infrared divergences or iterations of lower order contributions.  In the classical limit we are thus permitted to drop all recoil effects on the other particle, which can then be represented by a static background.  

\subsection{Lifting to Arbitrary Mass Ratio}

At leading PM order the test-particle limit encodes sufficient data to reconstruct the perturbed scattering amplitude 
for arbitrary mass ratio in an arbitrary reference frame.  To do this start we start with the test-particle amplitude and boost away from the rest frame by sending 
\eq{
E \;\; \rightarrow \;\; m_1\sigma  \qquad \textrm{ and }\qquad 
p \;\; \rightarrow \;\; m_1\sqrt{\sigma^2-1} \qquad \textrm{ where } \qquad \sigma = -\frac{P_1  \cdot P_2 }{m_1 m_2} \, .
}{eq:boost}
Here $P_{1} = (E_{1}, p_{1})$ and $P_{2} = (E_{2}, p_{2})$ are the four-momenta of the scattering bodies in a boosted frame where $E_{1,2}=\sqrt{p_{1,2}^2 + m_{1,2}^2}$.  At the same time, we also continue away from the test-particle limit to accommodate arbitrary masses $m_1$ and $m_2$.

Applying the replacement in \Eq{eq:boost}, we derive the leading PM tidal correction to the scattering amplitude for arbitrary mass ratio and in a general frame
\eq{
{\cal M}_1(P_1 , P_2,r) = \frac{m_1 m_2\sigma }{E_1 E_2}  {\cal M}_1(m_1 \sqrt{\sigma^2-1} , r)  \,, 
}{eq:boostM}
where the left-hand side of \Eq{eq:boostM} is the amplitude for arbitrary mass ratio in a general frame and the right-hand side is the amplitude in the test-particle limit in the rest frame of the heavy body, as defined in \Eq{eq:Mgen}.  Here the prefactor accounts for proper non-relativistic normalization after applying the replacement in \Eq{eq:boost}.   Note that if both bodies are tidally deformed then one should sum over \Eq{eq:boostM} with particle labels swapped.

This procedure similarly applies to the isotropic Hamiltonian, 
\eq{
H_1^{\rm iso}(P_1 , P_2 , r) = \frac{m_1 m_2\sigma }{E_1 E_2} H_1^{\rm iso}(m_1\sqrt{\sigma^2-1},r)\, ,
}{eq:boostH}   
where the left-hand side of \Eq{eq:boostH} is the Hamiltonian for arbitrary mass ratio in an arbitrary frame and the right-hand side is the isotropic gauge test-particle Hamiltonian in the rest frame of the heavy body.

While we have presented \Eq{eq:boostM} and \Eq{eq:boostH} in the context of tidal corrections, we emphasize that this basic procedure can be applied to any perturbation that deviates from the standard scenario of black holes interacting in general relativity.  

\subsection{Fourier Transform}

\Eq{eq:boostM} is an explicit formula for the leading PM scattering amplitude of tidally interacting bodies at arbitrary mass ratio written in terms of the amplitude in the test-particle limit.  According to \Eq{eq:M1} the latter
is essentially the isotropic gauge Hamiltonian in the test-particle limit.  Consequently, our only task is to map the test-particle Hamiltonian in \Eq{eq:H1_simp} to isotropic coordinates. Typically, this is achieved by explicitly constructing a coordinate transformation from $H_1(p,r,J)$ to $H_1^{\rm iso}(p,r)$, as we will do in \Sec{sec:all}. In this section, we derive a much simpler procedure applicable at the leading PM order.

Our basic idea is to take $H_1(p,r,J)$ and Fourier transform to momentum space to obtain $\widetilde H_1(p,q)$.  As discussed at length in \cite{2PM, 3PM, 3PMlong} one can compute the scattering amplitude mediated by the tidal corrections, treating $\widetilde H_1(p,q)$ as a Feynman vertex.  Crucially, at leading PM order, the contribution to the amplitude is exactly proportional to $\widetilde H_1(p,q)$ albeit evaluated at on-shell kinematics.  For a scattering particle with momentum $p$ and momentum transfer $q$, the on-shell condition implies that $p \cdot q \sim q^2$, which is subleading in classical counting and can thus be dropped.   From \Eq{eq:M1} we also know that the amplitude is given by the isotropic gauge Hamiltonian, so
\eq{
   \widetilde H_1(p,q) \bigg|_{p \cdot q =0} =\widetilde H_1^{\rm iso}(p,q) \, ,
}{}
where the replacement drops quantum contributions which are irrelevant to the classical dynamics.
We can then Fourier transform $\widetilde H_1^{\rm iso}(p,q)$ to obtain the isotropic Hamiltonian in position space, $H_1^{\rm iso}(p,r)$.

Conveniently, we can perform this procedure---Fourier transforming to momentum space, dropping terms that vanish on-shell, and Fourier transforming back to position space---in a single step.  This is encapsulated by the replacement,
\eq{\frac{J^{2k}}{r^n} \ \rightsquigarrow \ \frac{p^{2k} r^{2k}}{r^n} \times \frac{\textrm{Poch}\left(\frac{n}{2} -\frac{1}{2} -k, \ k\right)}{\textrm{Poch}\left(\frac{n}{2}-k, \ k\right)} \, ,
}{eq:FT}
where the left- and right-hand sides are exactly equal up to terms that Fourier transform to objects proportional to $p\cdot q$ which can be discarded on-shell.  See Appendix~\ref{app:FT} for details.  Also, note that this replacement must be used with care: it can only be applied after first expanding an expression fully in $J$ and $1/r$, since Fourier transform and multiplication do not commute. 

The upshot here is that \Eq{eq:FT} maps any non-isotropic Hamiltonian $H_1(p,r,J)$ to a physically equivalent isotropic Hamiltonian $H_1^{\rm iso}(p,r)$ at leading PM order.  After deriving  $H_1^{\rm iso}(p,r)$ we can then obtain the scattering amplitude via \Eq{eq:M1} and lift to arbitrary mass ratio via \Eq{eq:boostM} and \Eq{eq:boostH}.

\subsection{Tidal Operators}

Let us now utilize the tools derived above to compute the leading PM tidal corrections to the isotropic gauge Hamiltonian.   We emphasize that our procedure is purely algebraic and thus implemented with ease.   In particular, for a tidal operator given by a product of traces of matrices built from electric and magnetic Weyl, i.e.~${\cal O} =[ F_1({\cal E},{\cal B}) ] [ F_2({\cal E},{\cal B}) ] \cdots [ F_k({\cal E},{\cal B}) ] $, we simply calculate ${\cal O}$ using \Eq{eq:Emat} and \Eq{eq:Bmat} below, plug into \Eq{eq:H1_simp}, and apply the replacement rule in \Eq{eq:FT} to derive the isotropic Hamiltonian at arbitrary mass ratio in \Eq{eq:boostH}. As we will see, this procedure yields closed form results for certain infinite classes of tidal operators. We have verified that our results can also be obtained by the method of Ref.~\cite{Bini:2020flp}, and that we agree with all results collected in the Appendix of~\cite{Zvi_tidal} up to the choice of normalization. 

\subsubsection{Kretschmann Scalar to a Power: ${\cal O} = [C^2]^n$}

As a first trivial example let us consider a tidal operator given by the Kretschmann scalar to a power.  Computing $[C^2] = \frac{12 R^2}{r^6}$ in Schwarzschild coordinates and plugging into \Eq{eq:H1_simp}, we obtain the test-particle Hamiltonian
\eq{
H_1(p,r,J) &=H_1^{\rm iso}(p,r)= -\frac{1}{2E}\left( \frac{12 R^2}{r^6} \right)^n \, ,
}{}
which is automatically in isotropic form.  Boosting to an arbitrary frame via \Eq{eq:boostH}, we obtain the isotropic Hamiltonian at arbitrary mass ratio,
\eq{
H_1^{\rm iso}(P_1, P_2,r)= -\frac{m_2}{2 E_1 E_2}\left( \frac{48G^2 m_2^2}{r^6} \right)^n \,.
}{}
For $n=1$, this agrees with previous results~\cite{Bini:2020flp,PMGR,PMtidal,Kalin:2020lmz}. Note the relative factor of $1/2$ in the normalization of the tidal coefficient here as compared to~\cite{PMtidal}.

\subsubsection{Electric Weyl to a Power: ${\cal O} = [{\cal E}^n]$}

Next, consider a tidal operator given by the trace of a power of the electric Weyl tensor, 
\eq{
{\cal E}_{\alpha\beta} =\frac{1}{m^2} P^\mu  P^\nu C_{\mu\alpha\nu \beta} \, ,
}{}
where at leading order in tidal coefficients, $P$ is defined as in \Eq{eq:P} with an energy component equal to $H_0$. The choice of coordinates for the metric will not affect our answer at leading PM order since $C$ need only be evaluated on the Newtonian background.  The mixed index electric Weyl tensor at leading PM order is
\eq{ {\cal E}^\alpha_{\;\;\,\beta} &= \frac{R}{m^2 r^3} \times
\left(
\begin{array}{cccc}
 p^2-\frac{3 J^2}{2 r^2} & E p
   \sqrt{1-\frac{J^2}{p^2 r^2}} & 0 &
   -\frac{E J}{2} \\
 - E p \sqrt{1-\frac{J^2}{p^2 r^2}} &
  -E^2 -\frac{J^2}{2 r^2} & 0 & \frac{Jp}{2} \sqrt{1-\frac{J^2}{p^2 r^2}} \\
 0 & 0 & \frac{m^2}{2}+\frac{3 J^2}{2 r^2} & 0 \\
 \frac{E J}{2 r^2} & \frac{J p}{2 r^2} 
   \sqrt{1-\frac{J^2}{p^2 r^2}}& 0 &
  \frac{m^2}{2} + \frac{J^2}{2 r^2} \\
\end{array}
\right) \, ,
}{eq:Emat}
and its eigenvalues are
\eq{
\textrm{eig}\left[{\cal E}^\alpha_{\;\;\,\beta}\right] = \left\{ 0, \frac{ R}{2r^3},\frac{ R}{2r^3}\left(1+ \tfrac{3J^2}{m^2r^2  }\right)  ,-\frac{ R}{2r^3}\left(2+ \tfrac{3J^2}{m^2r^2  }\right) \right\} \, .
}{}
See App.~\ref{app:EB} for these expressions  at all orders in the PM expansion. Thus the electric Weyl tensor to the power $n$ is
\begin{widetext}
\eq{
{} [{\cal E}^n] = \textrm{Tr}({\cal E}{\cal E}\cdots {\cal E}) =  \left( \frac{ R}{2r^3}\right)^n \left[1 + \left(1+ \tfrac{3J^2}{m^2r^2  }\right)^n + (-1)^n\left(2+ \tfrac{3J^2}{m^2r^2  }\right)^n \right] \, .
}{}
\end{widetext}
We then apply a binomial expansion and then eliminate $J$ via the replacement in \Eq{eq:FT}.   Resumming terms analytically, we obtain the test-particle Hamiltonian in isotropic gauge,
\begin{widetext}
\eq{
H_1^{\rm iso}(p,r) &= -\frac{1}{2E}\left(\frac{ R}{2 r^3}\right)^n \left[ 1 + {}_2 F_1\left(-n,-\tfrac{1}{2} + \tfrac{3n}{2}, \tfrac{3n}{2},-\tfrac{3p^2}{m^2}\right) +(-2)^n {}_2 F_1\left(-n,-\tfrac{1}{2} + \tfrac{3n}{2}, \tfrac{3n}{2},-\tfrac{3p^2}{2m^2}\right)\right].
}{}
\end{widetext}
Applying \Eq{eq:boostH}, we derive a closed form expression for the isotropic Hamiltonian at arbitrary mass ratio, 
\eq{
H_1^{\rm iso}(P_1, P_2,r) &\overset{\phantom{n=1}}{=}  -\frac{m_2}{2E_1 E_2}\left(\frac{G  m_2}{ r^3}\right)^n \left[ 1 + {}_2 F_1\left(-n,-\tfrac{1}{2} + \tfrac{3n}{2}, \tfrac{3n}{2},-3(\sigma^2-1)\right) +(-2)^n {}_2 F_1\left(-n,-\tfrac{1}{2} + \tfrac{3n}{2}, \tfrac{3n}{2},-\tfrac{3}{2}(\sigma^2-1)\right)\right] \\
&\overset{n=1}{=} 0\\
&\overset{n=2}{=} -\frac{m_2}{2E_1 E_2} \times \frac{3 (G  m_2)^2 \left(35
   \sigma ^4-30 \sigma
   ^2+11\right)}{8 r^6}\\
&\overset{n=3}{=}  \phantom{-}\frac{m_2}{2E_1 E_2}\times\frac{6 (G  m_2)^3 \left(40
   \sigma ^4-36 \sigma
   ^2+7\right)}{11 r^9} \\
&\overset{n=4}{=} -\frac{m_2}{2E_1 E_2} \times \frac{9 (G m_2)^4
   \left(12155 \sigma ^8-22880
   \sigma ^6+18590 \sigma ^4-7304
   \sigma ^2+1231\right)}{896
   r^{12}} \\
   &\;\;\; \vdots
      }{eq:H_En}
The case of $n=2$ agrees with~\cite{Bini:2020flp,PMGR,PMtidal,Kalin:2020lmz} while $n=3$ agrees with~\cite{Bini:2020flp}. We can similarly compute arbitrary products of traces of electric Weyl, noting that due to the simple form of the eigenvalues a trace of any product can be reduced to products of $[{\cal E}^{2}]$ and $[{\cal E}^{3}]$.

\subsubsection{Magnetic Weyl to a Power: ${\cal O} = [{\cal B}^{n}]$}

We can repeat the same procedure for a tidal operator given by the trace of a power of magnetic Weyl, 
\eq{
{\cal B}_{\alpha\beta} = \frac{1}{m^2}  P^\mu  P^\nu \tilde C_{\mu\alpha\nu \beta} \, ,
}{}
where $\tilde C_{\mu\nu\rho\sigma} = \tfrac{1}{2} \epsilon_{\mu\nu\alpha\beta} C^{\alpha \beta}_{\;\;\;\;\rho\sigma}$ is the dual Weyl tensor.    The mixed index magnetic Weyl tensor is
\eq{{\cal B}^\alpha_{\;\;\,\beta} &=  \frac{R}{m^2 r^3} \times
\left(
\begin{array}{cccc}
 0 & 0 & -\frac{3J p}{2}  \sqrt{1-\frac{J^2}{p^2
   r^2}} & 0 \\
 0 & 0 & \frac{3 E J}{2} & 0 \\
 \frac{3 J p}{2 r^2} \sqrt{1-\frac{J^2}{p^2 r^2}}
   & \frac{3 E J}{2 r^2} & 0 & 0 \\
 0 & 0 & 0 & 0 \\
\end{array}
\right)
}{eq:Bmat}
and its eigenvalues are
\eq{
\textrm{eig}\left[{\cal B}^\alpha_{\;\;\,\beta}\right] = \left\{  0, 0, -\frac{3J R \sqrt{1+\tfrac{J^2}{m^2 r^2} } }{2mr^4},\frac{3J R \sqrt{1+\tfrac{J^2}{m^2 r^2} } }{2mr^4} \right\} \, ,
}{}
with expressions at all PM orders in App.~\ref{app:EB}. Any trace of an odd power of ${\cal B}$ will vanish.  The tidal operator is then
\eq{
{}[{\cal B}^{n}] = \textrm{Tr}({\cal B}{\cal B}\cdots {\cal B}) = 2 \left(\frac{9J^2 R^2 (1+\tfrac{J^2}{m^2 r^2} )}{4m^2r^{8}}\right)^{n/2} \, .
}{eq:B2n}
Expanding and replacing $J$ via \Eq{eq:FT}, we obtain
\eq{
H_1^{\rm iso}(p,r) &=-\frac{1}{E} \left(\frac{3 R p}{2 mr^3}\right)^{n} \frac{\Gamma(3n/2)\Gamma(-\tfrac{1}{2}+2n)}{\Gamma(2n)\Gamma(-\tfrac{1}{2}+\frac{3n}{2})} \times  {}_2 F_1\left(-\frac{n}{2},-\tfrac{1}{2} +2n ,2n,-\tfrac{p^2}{m^2}\right) \,.
}{}
Again applying \Eq{eq:boostH}, we obtain
\eq{
H_1^{\rm iso}(P_1, P_2,r) & \overset{\phantom{n=1}}{=} -\frac{m_2}{E_1 E_2} \left(\frac{3G m_2 \sqrt{\sigma^2-1}}{r^3}\right)^{n} \frac{\Gamma(3n/2)\Gamma(-\tfrac{1}{2}+2n)}{\Gamma(2n)\Gamma(-\tfrac{1}{2}+3n/2)} \times  {}_2 F_1\left(-n/2,-\tfrac{1}{2} +2n ,2n,-(\sigma^2-1)\right)\\
&\overset{n=2}{=} -\frac{m_2}{E_1 E_2} \times \frac{15 (G m_2)^2
   \left(\sigma ^2-1\right) \left(7
   \sigma ^2+1\right)}{16 r^6}\\
&\overset{n=4}{=} -\frac{m_2}{E_1 E_2}\times \frac{1287 (G m_2)^4
   \left(\sigma ^2-1\right)^2
   \left(85 \sigma ^4+10 \sigma
   ^2+1\right)}{1792 r^{12}}\\
   &\;\;\; \vdots
}{eq:H_Bn}
The case of $n=2$ agrees with previous results~\cite{Bini:2020flp,PMGR,PMtidal,Kalin:2020lmz}, and is consistent with the relation $[C^2]=8([{\cal E}^2]-[{\cal B}^2])$.  Due to the simple form of the eigenvalues of magnetic Weyl, any product of its traces can be reduced to a single trace $[{\cal B}^{n_1}][{\cal B}^{n_2}]\cdots [{\cal B}^{n_k}] = 2^{k-1} [{\cal B}^{n_1+n_2 +\cdots + n_k}]$.

\subsection{Beyond Schwarzschild}\label{sec:pert}

Our method for extracting kinematic data from the geodesic equation applies quite generally.  The only prerequisite for this approach is that the leading perturbative correction arises from a fan diagram in which the matter propagators are only on one side. For the remainder of this section we study scenarios in which the tidal interactions may be absent  but the system still has some small perturbative correction that deviates from a binary system of Schwarzschild black holes in general relativity.
 
In the examples below we will assume a static, spherically symmetric metric that is a function of some small perturbative parameter.  Plugging into \Eq{eq:H_compact}, we then expand to linear order in the perturbative coefficient. Applying the  replacement in \Eq{eq:FT}, we obtain the test-particle Hamiltonian in isotropic coordinates, from which we derive the associated scattering amplitude and Hamiltonian for arbitrary mass ratio via \Eq{eq:boostM} and \Eq{eq:boostH}.

\subsubsection{Higher Derivative Corrections}

As a concrete example let us consider a modification to general relativity given by the leading higher derivative correction to graviton self-interactions,
\eq{
\Delta S = \frac{\xi}{16\pi G}  \int d^4 x \; \sqrt{-g} \; R^{\mu\nu}_{\;\; \;\; \rho\sigma} R^{\rho\sigma}_{\;\; \;\; \alpha\beta} R^{\alpha\beta}_{\;\; \;\; \mu\nu} \,.
}{eq:R3} 
It is clear that at linear order in $\xi$ and leading PM order this correction contributions via a fan diagram which is accounted for by the geodesic equation.  In the presence of \Eq{eq:R3} the perturbed metric is \cite{Cai:2019npx}
\eq{
g_{tt}(r) = - \left( 1- \tfrac{R}{r} +\tfrac{5\xi R^3}{r^7}   \right) \,, \qquad  g_{rr}(r) = \left( 1- \tfrac{R}{r}+\tfrac{54\xi R^2}{r^6}  -\tfrac{49\xi R^3}{r^7}   \right)^{-1} \, , \qquad g_{\Omega}(r) =1 \, .
}{eq:metric_R3}
Next, we plug \Eq{eq:metric_R3} into \Eq{eq:H0} and expand $H = H_0 + \xi H_1$.   Solving to linear order in  $\xi$, we derive the perturbed Hamiltonian,
\eq{
H_1(p,r,J) = \frac{27 R^2\left(p^2 - \tfrac{J^2}{{r}^2}\right)}{E r^{6}}  \, ,
}{}
where we have truncated to leading PM order. Transforming to isotropic gauge via
 \Eq{eq:FT} yields
\eq{
H_1^{\rm iso}(p,r) = \frac{9 R^2 p^2 }{2E r^{6}} \, .
}{|}
We then boost to an arbitrary frame via \Eq{eq:boostH} and sum over the exchange of bodies 1 and 2 since \Eq{eq:R3} generates contributions from fan diagrams as well as their flipped partners.  We thus obtain
\eq{
H_1^{\rm iso}(P_1, P_2,r) = \frac{18 G^2 m_1^2 m_2^3 (\sigma^2-1)  }{E_1 E_2 r^{6}} + \{ 1\leftrightarrow 2\} \,,
}{}
which agrees with known results in the PN \cite{Brandhuber:2019qpg,Emond:2019crr} and PM \cite{Cristofoli:2019ewu, Zvi_tidal} expansions.

\subsubsection{Electric Charge}

Another application of our approach is the scattering of electrically charged bodies.  While this is obviously irrelevant to astrophysical black holes, we can still derive scattering amplitudes of more formal interest.  
To begin, we consider the interactions of a neutral body of mass $m_1=m$ and charge $q_1=0$ with an electrically charged body
of mass $m_2 = M$ and charge $q_2=Q$ in the test-particle limit, $m\ll M$.  We can define a charge-to-mass ratio parameter 
\eq{
z &=\frac{Q^2}{4 \pi G M^2} \, ,
}{eq:zdef} 
for which $0\leq z \leq 1$.  We take $z \ll 1 $ to be perturbatively small, i.e.~corresponding to a milli-charged body.

To compute the geodesic motion we use the Reissner-N\"{o}rdstrom metric in standard coordinates where
\eq{
g_{tt}(r) = - \left( { 1- \frac{R}{r} +\frac{z R^2}{4r^2} }   \right) \,, \qquad  g_{rr}(r) =  \left( { 1- \frac{R}{r} +\frac{z R^2}{4r^2} }   \right)^{-1}  \,, \qquad g_{\Omega}(r) = 1 \, .
}{eq:RN}
We then expand the Hamiltonian up to linear order in the charge-to-mass ratio, $H = H_0 + z H_1$, and solve the geodesic equation in \Eq{eq:geodesic} order by order in $z$.  At zeroth order in $z$, $H_0$ is obtained by inserting the metric for an uncharged black hole in Schwarzschild coordinates into \Eq{eq:H0}.   Meanwhile, at first order we obtain
\eq{
H_1(p,r,J) = \frac{R^2\left(m^2 +2 p^2 - \tfrac{J^2}{{r}^2}\right)}{8E r^{2}}  \,,
}{}
again truncating to leading PM order.  Applying the replacement rule in \Eq{eq:FT}, we trivially obtain the isotropic test-particle Hamiltonian
\eq{
H_1^{\rm iso}(p,r) = \frac{R^2\left(m^2 +\tfrac{3}{2} p^2 \right)}{8E r^{2}} \, .
}{eq:H1iso_RN}
Boosting to a general frame and lifting to arbitrary mass ratio we obtain
\eq{
H_1^{\rm iso}(P_1, P_2,r) =\ \frac{ G^2 m_1^2 m_2^3 (3\sigma^2-1)  }{4 E_1 E_2 r^{2}}  \, , 
}{eq:H1iso_RN_gen}
which describes a neutral body interacting with a charged body at lowest order in the charge.   The static limit of this expression agrees exactly with~\cite{BjerrumBohr:2002sx,Butt:2006gv,Faller:2007sy,Holstein:2008sy} after including the charge-to-mass ratio $z=\tfrac{q_2^2}{4 \pi G m_2^2}$ as defined in \Eq{eq:zdef}.   This result is trivially generalized to include multiple electric and magnetic charges simply by inserting the sum of all charges squared into $z$. 

We can verify \Eq{eq:H1iso_RN} by mapping to isotropic coordinates at the very beginning of the calculation.  In particular, we can apply the diffeomorphism 
\eq{
r \rightarrow f(r) = r + \frac{R}{2} + \frac{R^2}{16 r}(1-z) \, ,
}{}
which sends the metric components to
\eq{
g_{tt}^{\rm iso}(r) = - \left( \frac{ 1  -\frac{R^2}{16r^2} (1-z)}{ 1 + \frac{R}{2r} +\frac{R^2}{16r^2} (1-z)}   \right)^2 \,, \qquad  g_{rr}^{\rm iso}(r) =  g_{\Omega}^{\rm iso}(r)  = \left(  1 + \frac{R}{2r} +\frac{R^2}{16r^2} (1-z)   \right)^{2}   \, .
}{eq:RNiso}
Plugging into the geodesic equation and solving for $H^{\rm iso} = H_0^{\rm iso} + z H_1^{\rm iso}$ perturbatively in $z$ we obtain exactly \Eq{eq:H1iso_RN}.

\subsubsection{Electric Charge and Higher Derivative Corrections}

Last but not least, we consider the case of a neutral body interacting with a charged body in the presence of both tidal distortion and higher derivative corrections to the Einstein-Maxwell system,
\eq{
\Delta S  =& \phantom{{}+{}} \frac{\xi}{2} \int d^4 x \sqrt{-g}  \left( a_1 R^{\mu\nu} +a_2 g^{\mu\nu} R + a_3 F^{\mu\rho}F^{\nu}_{\,\,\,\, \rho} + a_4 g^{\mu\nu} F_{\rho\sigma} F^{\rho\sigma} \right) \nabla_\mu \phi \nabla_\nu\phi \\ 
& +  \xi \int d^4 x \sqrt{-g} \,  \big[ \, c_1 R^2 + c_2 R_{\mu\nu} R^{\mu\nu} +c_3 R_{\mu\nu\rho\sigma}R^{\mu\nu\rho\sigma} + c_4 R F_{\mu\nu} F^{\mu\nu} + c_5 R_{\mu\nu} F^{\mu\rho}F^{\nu}_{\,\,\,\, \rho} +c_6 R_{\mu\nu\rho\sigma}F^{\mu\nu}F^{\rho\sigma} \\
& \qquad \qquad \qquad \; +c_7 F_{\mu\nu}F^{\mu\nu}F_{\rho\sigma} F^{\rho\sigma} + c_8 F_{\mu\nu}F^{\nu\rho} F_{\rho\sigma}F^{\sigma\mu} \big] \, ,
}{eq:higherdim}
where the global prefactor $\xi$ characterizes the overall size of the perturbations.
Here we have included tidal corrections on the neutral body, so the geodesic equation is \Eq{eq:geodesic} where $\lambda =\xi$ and the tidal operator is
\eq{
{\cal O} &=  ( a_1 R^{\mu\nu} +a_2 g^{\mu\nu} R + a_3 F^{\mu\rho}F^{\nu}_{\,\,\,\, \rho} + a_4 g^{\mu\nu} F_{\rho\sigma} F^{\rho\sigma}) P_\mu P_\nu \, .
}{eq:Opert}
Of course, tidal interactions of this form can be eliminated with a field redefinition, but this fact will provide a useful consistency check of our final answer.
Finally, let us define natural  dimensionless coefficients,
\eq{
b_{1,2} = a_{1,2} , \qquad
b_{3,4} = (8\pi G )^{-1}  a_{3,4} , \qquad
d_{1,2,3} = (8\pi G )  c_{1,2,3} , \qquad
d_{4,5,6} =  c_{4,5,6} ,\qquad
d_{7,8} = (8\pi G )^{-1}  c_{7,8} \, ,
}{}
for later convenience.  

In this setup we consider the dynamics for arbitrary charge-to-mass ratio $z$ but treating the the overall scale of the higher derivative corrections $\xi$ as a small parameter.   Since the $d_i$ coefficients modify the extremality condition for large black holes they are intimately linked to the weak gravity conjecture~\cite{ArkaniHamed:2006dz}, which mandates the instability of such objects to avoid remnant pathologies.  Motivated by these connections, the linear in $d_i$ corrections to the Reissner-N\"{o}rdstrom metric were computed in \cite{Kats:2006xp}, and were later used to demonstrate the equivalence of the weak gravity conjecture to certain positivity conditions on black hole entropy \cite{Cheung:2018cwt,Cheung:2019cwi} that are valid in any tree-level ultraviolet completion of \Eq{eq:higherdim}.
The corrected metric from \cite{Kats:2006xp} is 
\eq{
g_{tt}(r) = - \left( { 1- \frac{R}{r} +\frac{z R^2}{4r^2} }  + \xi F(r)  \right) \,, \qquad  g_{rr}(r) =  \left( { 1- \frac{R}{r} +\frac{z R^2}{4r^2} + \xi G(r)}   \right)^{-1}  \,, \qquad g_{\Omega}(r) = 1 \, ,
}{eq:RN}
where the perturbation functions to leading PM order are
\eq{
F(r)&= - \frac{z R^2}{r^4} \left(d_2 + 4d_3 - 2d_4 +d_6\right) \\
G(r)&= - \frac{z R^2}{r^4} \left(2d_2 + 8d_3 +8d_4 +3d_5+ 4 d_6\right)   \, .
}{}
We again expand the Hamiltonian as $H = H_0 + \xi H_1$, this time inserting \Eq{eq:Opert} into \Eq{eq:geodesic} for 
$\lambda=\xi$ and solving order by order in $\xi$. At zeroth order in $\xi$ we obtain the Hamiltonian for a test-particle in the Reissner-N\"{o}rdstrom background while at linear order we find
\eq{
H_1(p,r,J) = \frac{z R^2}{E r^4} & \big[ \, m^2(-\tfrac{b_1}{8}-\tfrac{b_3}{4}-\tfrac{b_4}{2} -\tfrac{d_2}{2} -2 d_3 +d_4 -\tfrac{d_6}{2}) + p^2 (-\tfrac{3d_2}{2} -6d_3 -3d_4 -\tfrac{3d_5}{2} -\tfrac{5d_6}{2})  \\
&  \, + \tfrac{J^2}{r^2} (-\tfrac{b_1}{4}-\tfrac{b_3}{4} + d_2 +4d_3 +4d_4 +\tfrac{3d_5}{2} +2 d_6)\big] \, ,
}{}
at leading PM order.
Again applying \Eq{eq:FT}, we obtain the isotropic gauge Hamiltonian in the test-particle limit,
\eq{
H_1^{\rm iso}(p,r) = \frac{z R^2}{E r^4} \left[  m^2(-\tfrac{b_1}{8}-\tfrac{b_3}{4}-\tfrac{b_4}{2} -\tfrac{d_2}{2} -2 d_3 +d_4 -\tfrac{d_6}{2}) + p^2 (-\tfrac{3b_1}{16}-\tfrac{3b_3}{16} -\tfrac{3d_2}{4} -3d_3  -\tfrac{3d_5}{8} -d_6) \right] \, .
}{}
Lifting to arbitrary mass ratio and inserting $z=\tfrac{q_2^2}{4 \pi G m_2^2}$, we obtain 
\eq{
H_1^{\rm iso}(P_1, P_2,r) = \frac{ G m_1^2 m_2q_2^2}{4 \pi E_1 E_2 r^4} \left[   ( -\tfrac{3b_1}{4}-\tfrac{3b_3}{4} -3d_2 -12 d_3 -\tfrac{3d_5}{2}-4d_6 )\sigma^2 +(\tfrac{b_1}{4}-\tfrac{b_3}{4}  -2b_4  + d_2 + 4d_3 + 4 d_4 + \tfrac{3d_5}{2} +2 d_6) \right] \,,
}{eq:H1d}
which describes the dynamics of a neutral and charged body interacting via higher derivative corrections.  A similar exercise can be done in the presence of electric and magnetic charges using the perturbed metric computed in \cite{Cheung:2019cwi}.

A useful consistency check of \Eq{eq:H1d} comes from invariance of physical quantities under field redefinitions of the graviton in the effective field theory defined by \Eq{eq:higherdim}.  In particular, we consider a redefinition of the metric \cite{Cheung:2018cwt},
\eq{
g_{\mu\nu} \rightarrow g_{\mu\nu} +\delta g_{\mu\nu} \qquad {\rm where} \qquad  \delta g_{\mu\nu} = r_1 R_{\mu\nu} +r_2 g_{\mu\nu} R  + 8\pi G ( r_3  F_{\mu\rho}F_{\nu}^{\,\,\,\, \rho}  + r_4 g_{\mu\nu} F_{\rho\sigma} F^{\rho\sigma}) \, ,
}{}
which corresponds to a shift of the tidal coefficients by
\eq{
  b_1 &\rightarrow b_1 - r_1 \\
    b_2 &\rightarrow b_2 - r_2 \\
     b_3 &\rightarrow 
 b_3 - r_3 \\
 b_4 &\rightarrow b_4 - r_4 \, ,
 }{eq:bshift}
 and a shift of the higher dimension operator coefficients by
 \eq{
d_1 &\rightarrow d_1 - r_1/4 - r_2/2 \\
d_2 &\rightarrow d_2 + r_1/2\\
 d_3 &\rightarrow d_3 \\
  d_4 &\rightarrow d_4 + r_1/8 - r_3/4 - r_4/2 \\
   d_5 &\rightarrow d_5 - r_1/2 + r_3/2 \\
    d_6 &\rightarrow d_6 \\
     d_7 &\rightarrow 
 d_7 + r_3/8 \\
  d_8 &\rightarrow d_8 - r_3/2 \, .
}{eq:dshift}
We then find that \Eq{eq:H1d} is invariant under \Eq{eq:bshift} and \Eq{eq:dshift}, as expected since the isotropic Hamiltonian is proportional to the physical scattering amplitude at leading PM order.

\section{All Orders in $G$}\label{sec:all}

Test-particle dynamics can offer useful consistency checks for higher order PM calculations.  In this section we present a prescription for analytically deriving the isotropic gauge Hamiltonian in the test particle limit for a general tidal moment at all PM orders.  From this quantity we then derive closed form expressions for the corresponding scattering amplitudes at all PM orders.

\subsection{Diffeomorphism to Isotropic Coordinates}

To begin, we apply a radius-dependent diffeomorphism $r\rightarrow f(r)$ to an initial metric of the form in \Eq{eq:metric}.  The transformed line element is
\eq{
ds^2 = g_{tt}(f(r)) dt^2 + g_{rr}(f(r)) f'(r)^2 dr^2 + g_{\Omega}(f(r))  f(r)^2 d\Omega \,.
}{eq:metric2}
Any diffeomorphism of the metric will produce a new set of coordinates that automatically preserves the usual Poisson bracket structure of the corresponding phase space variables.  Crucially, $f(r)$ can be an implicit function of constants of motion such as the energy $H$ and the angular momentum $J$, so we define
\eq{
f(r) =  r + \lambda c(r,H, J) \,,  \qquad c(r,H, J) = \sum_{a=0}^\infty \sum_{b=0}^\infty c_{ab}(r)  H^{2a} J^{2b} \,.
}{eq:diff2} 
By construction, the nontrivial component of the diffeomorphism starts at linear order in the tidal coefficients so it only modifies tidal corrections to the Hamiltonian.  For our purposes $c(r,H,J)$ will be a polynomial and only a finite number of the $c_{ab}$ coefficients will be nonzero.  Note that since $H$ and $J$ are constants of motion, derivatives with respect to $r$ yield
$f'(r) = 1 + \lambda c'(r,H,J)$, where we define 
\eq{
c'(r,H,J) &= \sum_{a=0}^\infty \sum_{b=0}^\infty   c_{ab}'(r) H^{2a} J^{2b} \, ,
}{eq:cp}
i.e.~derivatives do not act on $H$ or $J$.

Earlier, we solved the geodesic equation in \Eq{eq:geodesic} to obtain the Hamiltonian at zeroth order in the tidal coefficients $H_0(p,r,J)$ in \Eq{eq:H0}.  Since the tidal operator enters algebraically as a mass deformation, we can solve \Eq{eq:geodesic} at linear order in the tidal coefficient to obtain
\begin{align}\label{eq:Hs2}
H(p,r,J)=& \ \sqrt{-g_{tt}(f(r))} \, \sqrt{ m^2  - \lambda  {\cal O}(p,r ,H_0, J)  + \tfrac{p^2 - \frac{J^2}{r^2}}{g_{rr}(f(r))f'(r)^2} + \tfrac{J^2}{g_\Omega(f(r)) f(r)^2}  }  + {\cal O}(\lambda^2) \,.
\end{align}
Noting the implicit $\lambda$ dependence in $f(r)$, we then decompose \Eq{eq:Hs2} as $H = H_0 + \lambda H_1$ and expand to linear order in $\lambda$.  Similar to before, any term entering with an explicit factor of $\lambda$ can be simplified since any appearance of $H$ can be replaced with the point-particle Hamiltonian in the absence of tidal effects, $H_0$.  So concretely, the tidal operator should be evaluated as ${\cal O}(p,r,H_0, J)$ and the diffeomorphism functions should be evaluated as $c(r,H_0,J)$ and $c'(r,H_0,J)$.

While $H_0(p,r,J)$ will in general have $J$ dependence we can simplify our calculation by using the isotropic coordinates in \Eq{eq:iso_coord} for the initial metric before applying the diffeomorphism.  We will assume this for the remainder of this section.
By contrast, the tidal Hamiltonian $H_1(p,r,J)$ has $J$ dependence even if the initial metric is in isotropic coordinates.  Since $H_1(p,r,J)$ is quite complicated we do not write it explicitly here, but the procedure for computing it is completely mechanical and described above.

\begin{figure}[t]
\begin{center}
\includegraphics[scale=.48]{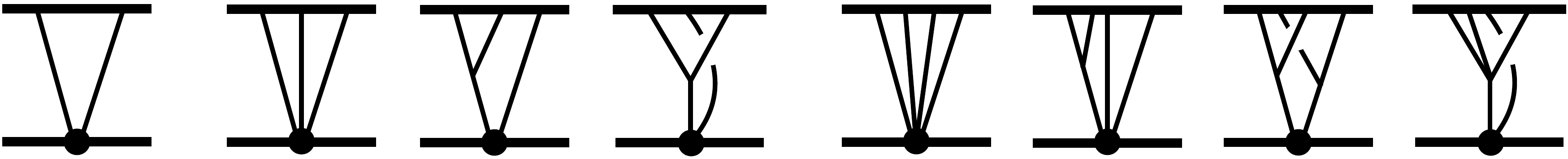}
\end{center}
\caption{To all orders in PM, contributions to the scattering amplitude of a tidally distorted test particle are generated by fan diagrams. Shown here are examples at one-, two-, and three-loop orders.}
\label{fig:fan_allPM}
\end{figure}

In the final step, we solve for the coefficients $c_{ab}(r)$ of the diffeomorphism to exactly cancel all $J$ dependence in $H_1(p,r,J)$. 
We immediately find that $H_1(p,r,J)$ is a polynomial in $J$ of finite degree, so only a finite number of terms are needed in the diffeomorphism in \Eq{eq:diff2}.  Demanding that the coefficient of every positive power of $J$ is zero then produces a set of differential equations for the coefficients $c_{ab}(r)$ in \Eq{eq:diff2}.  These equations can be solved analytically, yielding closed form expressions for $c_{ab}(r)$, which turn out to be rational functions of the radius $r$.\footnote{The constants of integration are fixed so that the coefficients $c_{ab}(r)$ do not blow up at large $r$.}  

Inserting the solved-for diffeomorphism back into $H_1(p,r,J)$ yields the tidal Hamiltonian in isotropic coordinates,
\begin{align}\label{eq:Hiso}
H^{\rm iso}_1(p,r) =&  \  \frac{g^{\rm iso}_{tt}(r)}{2 H^{\rm iso}_0}  {\cal O}(p,r,H^{\rm iso}_0,0) + \left[ {  H^{\rm iso}_0 {g^{\rm iso}_{tt}}'(r) \over 2 g^{\rm iso}_{tt}(r)} + {p^2 g^{\rm iso}_{tt}(r) {g^{\rm iso}_{rr}}'(r) \over 2 H^{\rm iso}_0 g^{\rm iso}_{rr}(r)^2 } \right] c(r,H_0^{\rm iso}, 0) + {p^2 g^{\rm iso}_{tt}(r) \over H^{\rm iso}_0 g^{\rm iso}_{rr}(r) } c'(r,H_0^{\rm iso},0) \,, 
\end{align}
where $c(r,E,J)$ and $c'(r,E,J)$ are defined in \Eq{eq:diff2} and \Eq{eq:cp}, and must be solved for to eliminate all $J$ dependence in $H_1(p,r,J)$.  Explicit expressions for these quantities will be presented in the examples given below.   We again emphasize that all components of the initial metric are evaluated in isotropic coordinates.  Importantly, every step of this procedure can be performed at all PM orders.

Following the procedure described in \Sec{sec:HtoA}, we obtain the scattering amplitude in the test-particle limit, 
\eq{
  {\cal M}_1(p,r) &=  \frac{1}{2E} \left[  g^{\rm iso}_{rr}(r)  {\cal O}(\sqrt{E^2 + 2E {\cal M}_0 - m^2},r,E, 0) -{g^{\rm iso}_{rr}}'(r) \left(m^2 +\tfrac{E^2}{g^{\rm iso}_{tt}(r)} \left(1 - \tfrac{g^{\rm iso}_{rr}(r){g^{\rm iso}_{tt}}'(r)}{g^{\rm iso}_{tt}(r) {g^{\rm iso}_{rr}}'(r)}  \right)\right) c(r,E,0) \right. \\
  & \quad \left. -2 g^{\rm iso}_{rr}(r) \left(m^2 +\tfrac{E^2}{g^{\rm iso}_{tt}(r)} \right) c'(r,E,0)\right] + \textrm{iteration} \,,
}{eq:Miso}
together with ${\cal M}_0(p,r)$ as defined in \Eq{eq:Mgen}.
These results are in the test-particle limit and at all orders in PM, and are equivalent to the resummation of the Feynman diagrams shown in \Fig{fig:fan_allPM}.

It is straightforward to compute the scattering angle to arbitrarily high PM order from either the Hamiltonian or the scattering amplitude (see for example~\cite{Vines:2018gqi, 3PMlong, B2B1, BDG2}). We define the scattering angle as $\chi = \chi_0 + \lambda \chi_1$, where $\chi_0$ arises from minimal gravitational coupling and $\chi_1$ is the tidal correction at linear order. For $\chi_0$ we find
 \eq{
  \chi_0 &= \frac{mR \left(2 
   u^2+1\right)}{J
   u}+\frac{3 \pi  m^2
   R^2 \left(5
   u^2+4\right)}{16
   J^2}+\frac{m^3 R^3 \left(64
   u^6+72
   u^4+12
   u^2-1\right)}{12 J^3
   u^3}+ {\cal O}(R^4) \,,
 }{}
 where $u =p_\infty/m$ and $p_\infty$ is the momentum at infinity. In all the examples we consider, we find that $\chi_1$  agrees when derived from $H_1(p,r,J)$ versus $H_1^{\rm iso}(p,r)$, showing they are gauge equivalent.

\subsection{Examples}

\begin{table}[t]
\setlength{\tabcolsep}{5pt} 
\renewcommand{\arraystretch}{3}
\begin{tabular}{|c|c|}
\hline
 $[{\cal E}^2] $ & $\frac{6144 \rho ^6 \left(768 J^4 \rho ^4+48 J^2 m^2 \rho ^2 (\rho +1)^4 R^2+m^4 (\rho +1)^8 R^4\right)}{ m^4 (\rho +1)^{20} R^8}$ \\
\hline
$[{\cal E}^3]$ & $ -\frac{196608 \rho ^9 \left(1152 J^4 m^2 \rho ^4+72 J^2 m^4 \rho ^2 (\rho +1)^4 R^2+m^6 (\rho +1)^8 R^4\right)}{m^6 (\rho +1)^{26} R^{10}}$ \\
\hline
$[{\cal E}^4]$ & $\frac{18874368 \rho ^{12} \left(768 J^4 \rho ^4+48 J^2 m^2 \rho ^2 (\rho +1)^4 R^2+m^4 (\rho +1)^8 R^4\right)^2}{m^8 (\rho +1)^{40} R^{16}}$ \\
\hline
$ [d{\cal E}^2]$ & $\frac{294912 (\rho -1)^2 \rho ^8 \left(2 m^2 \rho ^2 (\rho +1)^4 R^2 \left(96 J^2+p^2 \rho ^2 R^2\right)+128 \left(10 J^4 \rho ^4+J^2 p^2 \rho ^6 R^2\right)+5 m^4 (\rho +1)^8 R^4\right)}{m^4(\rho +1)^{26} R^{10}}$ \\
\hline
 $[{\cal B}^2] $ & $ \frac{294912 J^2 \rho ^8 \left(16 J^2 \rho ^2+m^2 (\rho +1)^4 R^2\right)}{m^4 (\rho +1)^{20} R^8}$ \\
\hline
\end{tabular}
\caption{Tidal operators evaluated on a background Schwarzschild metric in isotropic coordinates.  As discussed in the text, since we are working at linear order in the tidal coefficients we have set the energy component of the test-particle momentum to be the Hamiltonian $H_0$ in the absence of tidal corrections.}
\label{table:operators}
\end{table}

\begin{table}[t]
\setlength{\tabcolsep}{1pt} 
\renewcommand{\arraystretch}{3}
\begin{tabular}{|c|c|}
\hline
\tabeq{1cm}{
&[{\cal E}^2] \\
&\\
&[{\cal B}^2]
}
 &
\tabeq{16.5cm}{
c_{00} &= 
-\tfrac{16  \left(15015 \rho
   ^7+7865 \rho ^6+429 \rho ^5-2405
   \rho ^4-1755 \rho ^3-605 \rho
   ^2-105 \rho -7\right)}{5005 \rho
    (\rho +1)^{11} m^2 R^3} \\
c_{01} &= -\tfrac{512 \rho  \left(6435 \rho
   ^7+5005 \rho ^6+3003 \rho
   ^5+1365 \rho ^4+455 \rho ^3+105
   \rho ^2+15 \rho +1\right)}{715
   (\rho +1)^{15} m^4R^5} \\
   c_{10} &= 
-\tfrac{16 \left(105105 \rho
   ^7+43615 \rho ^6-9009 \rho
   ^5-17615 \rho ^4-6825 \rho
   ^3-695 \rho ^2+105 \rho
   +7\right)}{5005 (\rho -1)^2 \rho
    (\rho +1)^9 m^4R^3}
 }
    \\
\hline
$[{\cal E}^3]$ & 
\tabeq{16.5cm}{
c_{00} &= 
\tfrac{1536  \left(1175720 \rho
   ^{10}+764218 \rho ^9+239666 \rho
   ^8-98192 \rho ^7-180880 \rho
   ^6-124355 \rho ^5-53599 \rho
   ^4-15428 \rho ^3-2884 \rho
   ^2-315 \rho -15\right)}{1616615
   \rho  (\rho +1)^{17} m^2 R^5}   \\
c_{01} &= \tfrac{147456  \rho  \left(352716
   \rho ^{10}+293930 \rho ^9+203490
   \rho ^8+116280 \rho ^7+54264
   \rho ^6+20349 \rho ^5+5985 \rho
   ^4+1330 \rho ^3+210 \rho ^2+21
   \rho +1\right)}{323323 (\rho
   +1)^{21} m^4 R^7} \\
c_{10} &= \tfrac{1536  \left(11757200 \rho
   ^{10}+6995534 \rho ^9+1406342
   \rho ^8-1547816 \rho ^7-1718360
   \rho ^6-870485 \rho ^5-253897
   \rho ^4-38836 \rho ^3-1372 \rho
   ^2+315 \rho +15\right)}{1616615
   (\rho -1)^2 \rho  (\rho +1)^{15}
   m^4R^5} 
   }  
\\
\hline
$ [d{\cal E}^2]$ &
\tabeq{17.5cm}{
 c_{00} &= 
\tfrac{768  \left(4849845 \rho
   ^{11}-4555915 \rho ^{10}+352716
   \rho ^9+388892 \rho ^8+339796
   \rho ^7+226100 \rho ^6+113050
   \rho ^5+41762 \rho ^4+11039 \rho
   ^3+1967 \rho ^2+210 \rho
   +10\right)}{1616615 \rho  (\rho
   +1)^{17} m^2R^5} \\ 
 c_{01} &= -\tfrac{49152 \rho  \left(1939938
   \rho ^{11}-1587222 \rho
   ^{10}+293930 \rho ^9+203490 \rho
   ^8+116280 \rho ^7+54264 \rho
   ^6+20349 \rho ^5+5985 \rho
   ^4+1330 \rho ^3+210 \rho ^2+21
   \rho +1\right)}{323323 (\rho
   +1)^{21} m^4 R^7} \\
    c_{10} &= -\tfrac{768 \left(121246125 \rho
   ^{11}-140351575 \rho
   ^{10}-6113744 \rho ^9+7795928
   \rho ^8+12684856 \rho ^7+9405760
   \rho ^6+4318510 \rho ^5+1262702
   \rho ^4+219051 \rho ^3+17927
   \rho ^2+210 \rho
   +10\right)}{1616615 (\rho -1)^2
   \rho  (\rho +1)^{15} m^4 R^5}
   }
   \\
\hline
\end{tabular}
\caption{Coefficients specifying the diffeomorphism in \Eq{eq:cp} that goes to isotropic coordinates. }
\label{table:coeffs}

\vspace{1.5cm}

\setlength{\tabcolsep}{1pt} 
\renewcommand{\arraystretch}{3}
\begin{tabular}{|c|p{0.94\textwidth}|}
\hline
$[{\cal E}^2]$ & 
\scalebox{0.74}{\tabeq{15cm}{
&\frac{- 64}{5005 m^4 (\rho +1)^{22} R^4 H_0^{\rm iso}} \bigg[ 520 m^4 (\rho -1) \rho  \big(462 \rho ^6+198 \rho ^4-33 \rho ^3-77 \rho ^2-33 \rho -5\big) (\rho +1)^8 +2 m^2 p^2 (\rho -1) \rho ^3 \big(300300 \rho ^8  +49335 \rho ^7+160875 \rho ^6+143 \rho ^5 -29965 \rho ^4 \\
& -6955 \rho ^3+2465 \rho ^2+965 \rho  +21\big) (\rho +1)^4+p^4 (\rho -1) \rho ^6 \big(525525 \rho ^9 -53625 \rho ^8+226512 \rho ^7+16952 \rho ^6-26598 \rho ^5-4090 \rho ^4+3240 \rho ^3 +464 \rho ^2-231 \rho -21\big) \bigg]}
}  \\
\hline
$[{\cal E}^3]$ & 
\scalebox{0.74}{\tabeq{15cm}{
&\frac{6144}{1616615 m^4 (\rho +1)^{28} R^6H_0^{\rm iso}} \bigg[ 2128 m^4 (\rho -1) \rho  \big(12155 \rho ^9+7293 \rho ^7+1547 \rho ^6-1547 \rho ^5-1785 \rho ^4-935 \rho ^3-289 \rho ^2-51 \rho -4\big) (\rho +1)^8+2 m^2 p^2 (\rho -1) \rho ^3 \big(51731680 \rho ^{11} \\
&-13873496 \rho ^{10}+24043474 \rho ^9+7746186 \rho ^8-1224816 \rho ^7-2568496 \rho ^6-1146327 \rho ^5-164787 \rho ^4+48412 \rho ^3+23548 \rho ^2+3073 \rho +45\big) (\rho +1)^4 \\
&+p^4 (\rho -1) \rho ^6 \big(94057600 \rho ^{12}-42854994 \rho ^{11}+33404014 \rho ^{10}+12290150 \rho ^9+341734 \rho ^8-2042329 \rho ^7-875273 \rho ^6-38437 \rho ^5+73675 \rho ^4+17703 \rho ^3-1193 \rho ^2-765 \rho \\
& -45\big) \bigg]}
}
\\
\hline
$ [d{\cal E}^2]$ & 
\scalebox{0.74}{\tabeq{15cm}{
&\frac{-3072}{1616615 m^4 (\rho +1)^{28} R^6 H_0^{\rm iso}} \bigg[ 1064 m^4 (\rho -1) \rho  \big(364650 \rho ^{10}-875160 \rho ^9+838695 \rho ^8-376805 \rho ^7+13923 \rho ^6 +23205 \rho ^5 +17255 \rho ^4+7905 \rho ^3+2295 \rho ^2+391 \rho \\
&+30\big) (\rho +1)^8+2 m^2 p^2 (\rho -1) \rho ^3 \big(484984500 \rho ^{12} -1168812645 \rho ^{11}+1130895675 \rho ^{10} -492156392 \rho ^9+23740500 \rho ^8+31407228 \rho ^7+20575100 \rho ^6+7936110 \rho ^5 \\
& +1617546 \rho ^4+43225 \rho ^3-49959 \rho ^2-7350 \rho +30\big) (\rho +1)^4+p^4 (\rho -1) \rho ^6 \big(848722875 \rho ^{13}-2214027725 \rho ^{12}  +2178109479 \rho ^{11}-849478049 \rho ^{10}+32687600 \rho ^9 \\
& +37685056 \rho ^8 +22228214 \rho ^7+7028518 \rho ^6+558467 \rho ^5 -421925 \rho ^4-157773 \rho ^3-19637 \rho ^2-510 \rho -30\big) \bigg] }
}
\\
\hline
$[{\cal B}^2]$ & 
\scalebox{0.74}{\tabeq{15cm}{
&\frac{-64}{5005 m^4 (\rho +1)^{22} R^4 H_0^{\rm iso}} \bigg[ 520 m^4 (\rho -1) \rho  \big(462 \rho ^5+198 \rho ^4-33 \rho ^3-77 \rho ^2-33 \rho -5\big) (\rho +1)^8  +2 m^2 p^2 (\rho -1) \rho ^3 \big(300300 \rho ^8 +49335 \rho ^7+160875 \rho ^6+143 \rho ^5-29965 \rho ^4 \\
&-6955 \rho ^3+2465 \rho ^2+965 \rho   +21\big) (\rho +1)^4+p^4 (\rho -1) \rho ^6 \big(525525 \rho ^9-53625 \rho ^8+226512 \rho ^7+16952 \rho ^6-26598 \rho ^5-4090 \rho ^4+3240 \rho ^3  +464 \rho ^2-231 \rho -21\big) \bigg]}
}
\\
\hline
\end{tabular}
\caption{Contribution to the isotropic Hamiltonian $H_1$ from tidal operators at all PM orders. }
\label{table:Hamiltonians}
\end{table}

\begin{table}
\setlength{\tabcolsep}{1pt} 
\renewcommand{\arraystretch}{3}
\begin{tabular}{|c|p{0.907\textwidth}|}
\hline
$[{\cal E}^2]$ & 
\scalebox{0.8}{\tabeq{15cm}{
&\frac{64}{5005 E m^4 (\rho -1)^5 \rho ^6 (\rho +1)^8 R^4} \bigg[ E^4 (\rho +1)^4 \big(525525 \rho ^9-53625 \rho ^8+226512 \rho ^7+16952 \rho ^6-26598 \rho ^5-4090 \rho ^4 +3240 \rho ^3+464 \rho ^2 -231 \rho -21\big) \\
&+m^4 \big(165165 \rho ^8+12870 \rho ^7+20592 \rho ^6+20098 \rho ^5+13390 \rho ^4+6050 \rho ^3  +1760 \rho ^2+294 \rho +21\big) (\rho -1)^5 -2 E^2 m^2 (\rho +1)^2 \big(225225 \rho ^9-102960 \rho ^8 \\
&+65637 \rho ^7+16809 \rho ^6+3367 \rho ^5 +2865 \rho ^4+775 \rho ^3-501 \rho ^2-252 \rho -21\big) (\rho -1)^2 \bigg]} } \\
\hline
$[{\cal E}^3]$ & 
\scalebox{0.8}{\tabeq{15cm}{
&\frac{-6144}{1616615 E m^4 (\rho -1)^5 \rho ^6 (\rho +1)^{14} R^6} \bigg[  E^4 (\rho +1)^4 \big(94057600 \rho ^{12}-42854994 \rho ^{11}+33404014 \rho ^{10}+12290150 \rho ^9 +341734 \rho ^8 -2042329 \rho ^7-875273 \rho ^6 \\
&-38437 \rho ^5+73675 \rho ^4+17703 \rho ^3-1193 \rho ^2-765 \rho -45\big) +m^4 \big(16460080 \rho ^{11}+1352078 \rho ^{10} +2188648 \rho ^9+2278442 \rho ^8+1777792 \rho ^7+1073975 \rho ^6 \\
&+501676 \rho ^5 +177821 \rho ^4+46144 \rho ^3+8239 \rho ^2+900 \rho +45\big) (\rho -1)^5 -2 E^2 m^2 (\rho +1)^2 \big(42325920 \rho ^{12}-28981498 \rho ^{11}  +9360540 \rho ^{10}+4543964 \rho ^9 \\
&+1566550 \rho ^8+526167 \rho ^7+271054 \rho ^6+126350 \rho ^5 +25263 \rho ^4-5845 \rho ^3-4266 \rho ^2  -810 \rho -45\big) (\rho -1)^2 \bigg]}
}
\\
\hline
$ [d{\cal E}^2]$ & 
\scalebox{0.8}{\tabeq{15cm}{
&\frac{3072}{1616615 E m^4 (\rho -1)^5 \rho ^6 (\rho +1)^{14} R^6} \bigg[ E^4 (\rho +1)^4 \big(848722875 \rho ^{13}-2214027725 \rho ^{12}+2178109479 \rho ^{11} -849478049 \rho ^{10} +32687600 \rho ^9+37685056 \rho ^8 \\
&+22228214 \rho ^7+7028518 \rho ^6+558467 \rho ^5-421925 \rho ^4-157773 \rho ^3  -19637 \rho ^2-510 \rho -30\big) +m^4 \big(266741475 \rho ^{11}-274089725 \rho ^{10}-6231316 \rho ^9 \\
&-4458692 \rho ^8-2665396 \rho ^7  -1311380 \rho ^6-520030 \rho ^5-161462 \rho ^4 -37639 \rho ^3-6167 \rho ^2-630 \rho -30\big) (\rho -1)^6 -2 E^2 m^2 (\rho +1)^2 \big(363738375 \rho ^{13} \\
&-1045215080 \rho ^{12}+1047213804 \rho ^{11}-357321657 \rho ^{10} +8947100 \rho ^9 +6277828 \rho ^8+1653114 \rho ^7-907592 \rho ^6-1059079 \rho ^5-465150 \rho ^4-107814 \rho ^3 \\
& -12287 \rho ^2-540 \rho -30\big) (\rho -1)^2 \bigg] }
}
\\
\hline
$[{\cal B}^2]$ &  
\scalebox{0.8}{\tabeq{15cm}{
&\frac{64}{5005 E m^4 (\rho -1)^5 \rho ^6 (\rho +1)^8 R^4} \bigg[ E^4 (\rho +1)^4 \big(525525 \rho ^9-53625 \rho ^8+226512 \rho ^7+16952 \rho ^6-26598 \rho ^5-4090 \rho ^4 +3240 \rho ^3+464 \rho ^2 -231 \rho -21\big) \\
&-m^4 \big(75075 \rho ^7+62205 \rho ^6+41613 \rho ^5+21515 \rho ^4+8125 \rho ^3+2075 \rho ^2+315 \rho +21\big) (\rho -1)^6 -2 E^2 m^2 (\rho +1)^2 \big(225225 \rho ^9-102960 \rho ^8+65637 \rho ^7 \\
& +16809 \rho ^6+3367 \rho ^5+2865 \rho ^4+775 \rho ^3  -501 \rho ^2-252 \rho -21\big) (\rho -1)^2 \bigg]}
}
\\
\hline
\end{tabular}
\caption{Contribution to the scattering amplitude ${\cal M}_1$ from tidal operators at all PM orders.  }
\label{table:amplitudes}

\vspace{1.5cm}

\setlength{\tabcolsep}{6pt} 
\renewcommand{\arraystretch}{3}
\begin{tabular}{|c|c|}
\hline
$[{\cal E}^2]$ & 
\tabeq{15cm}{
&\tfrac{45 \pi  u^4 \left(35 u^4+40 u^2+16\right) m^4 R^2}{512 J^6}+\tfrac{12 u^3 \left(160
   u^6+288 u^4+168 u^2+35\right) m^5 R^3}{35 J^7} +\tfrac{63 \pi  u^2 \left(9009 u^8+20790
   u^6+16800 u^4+5600 u^2+640\right) m^6 R^4}{8192 J^8} \\
   & +\tfrac{4  \left(14336 u^{11} +39424 u^9
   +40128 u^7 +18480 u^5 +3696 u^3 +231 u \right) m^7 R^5}{77 J^9} + {\cal O}(R^6) \\
} \\ 
\hline
$[{\cal E}^3]$ & 
\tabeq{15cm}{
 &-\tfrac{96 u^7 \left(40 u^4+44 u^2+11\right) m^7 R^3}{385 J^9}-\tfrac{189 \pi  u^6 \left(3861
   u^6+7128 u^4+3960 u^2+640\right) m^8 R^4}{32768 J^{10}} \\
   &-\tfrac{8 u^5 \left(40320 u^8+98280
   u^6+82940 u^4+27885 u^2+3003\right) m^9 R^5}{1001 J^{11}}  + {\cal O}(R^6)\\
   } \\
\hline
$[{\cal E}^4]$ & \tabeq{15cm}{
&\tfrac{891 \pi  u^{10} \left(12155 u^8+25740 u^6+22880 u^4+9856 u^2+1792\right) m^{10}
   R^4}{1048576 J^{12}} \\
   &+\tfrac{192 u^9 \left(2903040 u^{10}+8273664 u^8+9550464 u^6+5697720
   u^4+1763580 u^2+230945\right) m^{11} R^5}{1616615 J^{13}} + {\cal O}(R^6)\\
}\\
\hline
$ [d{\cal E}^2]$ & 
\tabeq{15cm}{
 &\tfrac{1575 \pi  u^6 \left(35 u^4+40 u^2+16\right) m^6 R^2}{1024 J^8}+\tfrac{96 u^5 \left(3680
   u^6+7392 u^4+4752 u^2+1155\right) m^7 R^3}{385 J^9} \\
   & +\tfrac{1701 \pi  u^4 \left(21593 u^8+56760
   u^6+53040 u^4+21120 u^2+3200\right) m^8 R^4}{32768 J^{10}} \\
   &+\tfrac{8 u^3 \left(1469440
   u^{10}+4615520 u^8+5479760 u^6+3045900 u^4+786786 u^2+75075\right) m^9 R^5}{1001
   J^{11}} + {\cal O}(R^6) \\
   } \\
\hline
$[{\cal B}^2]$ & 
\tabeq{15cm}{
&\tfrac{225 \pi  u^6 \left(7 u^2+8\right) m^4 R^2}{512 J^6}+\tfrac{24 u^5 \left(80 u^4+144
   u^2+63\right) m^5 R^3}{35 J^7}+\tfrac{1323 \pi  u^4 R^4 \left(429 u^6 m^6+990 u^4 m^6+720 u^2
   m^6+160 m^6\right)}{8192 J^8} \\
   &+\tfrac{32 u^3 \left(1792 u^8+4928 u^6+4752 u^4+1848 u^2+231\right)
   m^7 R^5}{77 J^9} +{\cal O}(R^6)\\
 }
   \\
\hline
\end{tabular}
\caption{Contribution to the scattering angle $\chi_1$ from tidal operators at several PM orders.}
\label{table:angles}
\end{table}

Next, let us apply the method just described to the following tidal moment operators:
\begin{align}
[{\cal E}^2] \,,[ {\cal E}^3]  \,, [{\cal E}^4] \,, [d{\cal E}^2] \,,  [{\cal B}^2]\, .
\end{align}
Here we have defined
$[d{\cal E}^2] = d{\cal E}_{\alpha\beta\gamma} d{\cal E}^{\alpha\beta\gamma} $ where
$d{\cal E}_{\alpha\beta\gamma}  =\tfrac{1}{m^2} P^\mu  P^\nu \nabla_\gamma C_{\mu\alpha\nu \beta} $.
To begin, we evaluate these operators on a background Schwarzschild metric in isotropic coordinates.  See App.~\ref{app:EB} for explicit expressions for electric and magnetic Weyl tensors at all PM orders.  As noted earlier, since we are working at linear order in the tidal coefficients we can insert $H_0$ for the time component of the four-momentum that defines electric and magnetic Weyl.  Our results in terms of $\rho = 4r /R$ are shown in Table.~\ref{table:operators}. 

As discussed, by setting all $J$ dependence in $H_1(p, r, J)$ to zero we derive a system of differential equations which can be solved to obtain the $c_{ab}(r)$ coefficients in Table.~\ref{table:coeffs}, where all coefficients not shown are vanishing. The solutions for $[{\cal E}^4]$ are too cumbersome to display here but are included in the ancillary file. Note that the solutions for $[{\cal E}^2]$ and $[{\cal B}^2]$ are the same since the difference of these operators is the Kretschmann scalar, which is independent of $J$.

From our results in Table.~\ref{table:operators} and Table.~\ref{table:coeffs}, we assemble the isotropic test-particle Hamiltonian from \Eq{eq:Hiso} as well as the scattering amplitude from \Eq{eq:Miso}.  The tidal corrections to the isotropic Hamiltonian and the scattering amplitude (modulo iterations) are summarized in Table.~\ref{table:Hamiltonians} and Table.~\ref{table:amplitudes}.
We emphasize again that these results are valid at all PM orders in the test-particle limit.  However, we have verified that our expressions for $[{\cal E}^2]$ and $[{\cal B}^2]$ are consistent with the 3PM results of \cite{PMtidal} in the test-particle limit.

Last but not least, as a check of our results we compute the tidal corrections to the scattering angle at several PM orders for both $H_1(p,r,J)$ and $H_1^{\rm iso}(p,r)$ and find that they agree, thus establishing their physical equivalence at that order. These scattering angles are presented in Table.~\ref{table:angles}. The Hamiltonian, amplitude, and scattering angle for $[{\cal E}^4]$ are included in the ancillary file.

\section{Conclusions}\label{sec:conclusions}
Geodesic motion encodes all the kinematic data needed to reconstruct any scattering process mediated by fan diagrams, i.e., topologies in which only propagators of one of the bodies is present.   In this work we have exploited this fact to derive the complete conservative dynamics---in the form of amplitudes and isotropic Hamiltonians---at leading PM order in various scenarios which deviate perturbatively from the minimal setup of a black hole binary system in general relativity.   As we have demonstrated, our approach only entails simple algebraic manipulations and can be applied to a wide range of examples such as tidal operators and higher derivative corrections to gravitational or electromagnetic interactions. Furthermore, we have derived a method for computing the test-particle scattering amplitude to all PM orders, which could offer a useful check of higher PM calculations.

\bigskip 
\noindent {\it Acknowledgements.} We thank Andreas Helset and Jan Steinhoff for comments on the manuscript. We thank Zvi Bern, Julio Parra-Martinez, Radu Roiban, Eric Sawyer and Chia-Hsien Shen for helpful discussions, especially regarding their concurrent work~\cite{Zvi_tidal}. C.C. and N.S. are supported by the DOE under grant no.~DE-SC0011632 and by the Walter Burke Institute for Theoretical Physics. M.P.S. is supported by the Mani L. Bhaumik Institute for Theoretical Physics and David Saxon Presidential Term Chair in Physics. We used \texttt{Mathematica}~\cite{Mathematica} in combination with \texttt{xAct}~\cite{xAct}.

\appendix

\section{Derivation of \Eq{eq:FT}}\label{app:FT}

The replacement in \Eq{eq:FT} eliminates all dependence on the angular momentum $J$ in favor of powers of $p$ and $r$.  Its derivation is straightforward. To begin, consider the expression
\begin{equation}
    \label{expr}
    \frac{(p \cdot r)^{2k} + z(n,k) \  p^{2k} \ r^{2k}}{r^n} \, ,
\end{equation}
where $z(n,k)$ is precisely chosen so that the Fourier transform of \Eq{expr} is zero at leading order in the classical limit. Our aim will be to derive $z(n,k)$.  

The Fourier transform of the tensor $(r_{i_1} r_{i_2} \cdots r_{i_{2k}})/r^{n}$ is trivially obtained by taking derivatives of the Fourier transform of the scalar $1/r^m$,
\eq{
    \frac{r_{i_1} r_{i_2} \cdots r_{i_{2k}}}{r^{n}} 
   = & \frac{(n-4k-2)!!}{(n-2)!!} \left[ \partial_{i_1} \partial_{i_2} \cdots \partial_{i_{2k}}\frac{1}{r^{(n-4k)}} \right. \\
&  - \sum_{j=0}^{k-1} (-1)^{j+k} \left.\frac{(n-2k+2j-2)!!}{(n-4k-2)!! \ r^{(n-2k+2j)}}\left( r_{i_1} r_{i_2}\cdots r_{i_{2j}} \delta_{i_{(2j+1)} i_{(2j+2)}} \delta_{i_{(2j+3)} i_{(2j+4)}}\cdots \delta_{i_{(2k-2)} i_{2k}} + \textrm{perm} \right) \right] \, ,
}{rrrn}
where the number of permutations in each summand is $\frac{(2k)!}{(2j)! \  (k-j)! \ 2^{(k-j)}}$.
Truncating all but leading order classical contribution of  \Eq{expr} effectively sets $(p \cdot q)$ to zero.  In the Fourier transform of \Eq{rrrn}, each derivative brings down a factor of $q$. Therefore, the first term vanishes when all free indices are contracted into the momentum $p$. Inserting \Eq{expr} we find that
\eq{
    z(n,k) =&   \frac{1}{(n-2)!!} \left( \sum_{j=1}^{k-1} (-1)^{j+k} \frac{(n-2k+2j-2)!! \ (2k)!}{(2j)! \  (k-j)! \ 2^{(k-j)}} z(n-2k-2j, j)  - \frac{2k! (-1)^{(k+1)} (n-2k-2)!! }{2^k \ k!} \right) \, .
}{recur}
The solution to this equation is
\begin{equation}
\label{fn}
    z(n,k) = -  \frac{(2k-1)!! \ (n-2k-2)!!}{(n-2)!!} \, ,
\end{equation}
which is shown simply by plugging into both sides of \Eq{recur}.  Starting from the left-hand-side of \Eq{eq:FT} and eliminating all factors of $(p\cdot r)$ via \Eq{expr} given \Eq{fn}, we obtain
\begin{equation}
\frac{J^{2k}}{r^n} = \sum_{j=0}^{2k} {k \choose j} \frac{(-1)^j \ (p \cdot r)^{2j} \ p^{2(k-j)}}{r^{(n-2k+2j)}} \rightsquigarrow \ \frac{p^{2k} r^{2k}}{r^n} \times \frac{\textrm{Poch}\left(\frac{n}{2} -\frac{1}{2} -k, \ k\right)}{\textrm{Poch}\left(\frac{n}{2}-k, \ k\right)} \, .
\end{equation}
which correctly reproduces the right-hand-side of \Eq{eq:FT}.

\section{Electric and Magnetic Weyl Tensors}\label{app:EB}

Here we summarize the electric and magnetic Weyl tensors in isotropic coordinates at all orders in the PM expansion.
The expressions below are written in terms of $f_\pm = 1 \pm {R \over 4r}$. The electric Weyl tensor is
\eq{ {\cal E}^\alpha_{\;\;\,\beta} &= \frac{R}{m^2 r^3} \times
\left(
\begin{array}{cccc}
 { (p^2-\frac{3 J^2}{2 r^2}) \over f_+^{10} } & { p
   \sqrt{1-\frac{J^2}{p^2 r^2}} \sqrt{p^2 + f_+^4 m^2 }  \over f_+^7 f_- } & 0 &
   -\frac{J \sqrt{p^2 + f_+^4 m^2 }}{2 f_+^7 f_-} \\
 - { p \sqrt{1-\frac{J^2}{p^2 r^2}} \sqrt{p^2 + f_+^4 m^2} f_- \over f_+^{13}} &
  -{ J^2 + 2r^2 (p^2 + f_+^4 m^2) \over 2 r^2 f_+^{10}} & 0 & \frac{Jp}{2 f_+^{10}} \sqrt{1-\frac{J^2}{p^2 r^2}}   \\
 0 & 0 & \frac{m^2}{2 f_+^6}+\frac{3 J^2}{2 r^2 f_+^{10} } & 0 \\
 \frac{J \sqrt{p^2 + f_+^4 m^2} f_-}{2 r^2 f_+^{13} } & \frac{J p}{2 r^2 f_+^{10}} 
   \sqrt{1-\frac{J^2}{p^2 r^2}}& 0 &
  \frac{m^2}{2 f_+^6} + \frac{J^2}{2 r^2 f_+^{10}} \\
\end{array}
\right) \, ,
}{eq:EmatPM}
and its eigenvalues are
\eq{
\textrm{eig}\left[{\cal E}^\alpha_{\;\;\,\beta}\right] = \left\{ 0, \frac{ R}{2r^3 f_+^6 },{R (3J^2 + f_+^4 m^2 r^2) \over 2m^2 r^5 f_+^{10} } , - {R (3J^2 +2 f_+^4 m^2 r^2) \over 2m^2 r^5 f_+^{10} }  \right\} \, .
}{}
The magnetic Weyl tensor is
\eq{{\cal B}^\alpha_{\;\;\,\beta} &=  \frac{R}{m^2 r^3} \times
\left(
\begin{array}{cccc}
 0 & 0 & - \frac{3J p}{2 f_- f_+^7}  \sqrt{1-\frac{J^2}{p^2
   r^2}} & 0 \\
 0 & 0 & \frac{3 J \sqrt{p^2 +  f_+^4 m^2}}{2 f_+^{10}} & 0 \\
 \frac{3 J p f_-}{2 r^2 f_+^{13} } \sqrt{1-\frac{J^2}{p^2 r^2}}
   & \frac{3 J \sqrt{p^2 + f_+^4 m^2}}{2 r^2 f_+^{10}} & 0 & 0 \\
 0 & 0 & 0 & 0 \\
\end{array}
\right) \, ,
}{eq:BmatPM}
and its eigenvalues are
\eq{
\textrm{eig}\left[{\cal B}^\alpha_{\;\;\,\beta}\right] = \left\{  0, 0, -\frac{3J R \sqrt{J^2+ f_+^4 m^2 r^2} }{2m^2r^5 f_+^{10} },\frac{3J R \sqrt{J^2+ f_+^4 m^2 r^2} }{2m^2r^5 f_+^{10} } \right\} \, .
}{}

\end{document}